\documentclass[aps,prb,preprint,showkeys]{revtex4-2}
\pdfoutput=1
\usepackage{amsmath}
\usepackage{amssymb}
\usepackage{amsfonts}
\usepackage{graphicx}
\usepackage{bm}
\usepackage{siunitx}
\usepackage{color}
\usepackage{hyperref}
\hypersetup{%
    pdfborder = {0 0 0}
}

\allowdisplaybreaks

\renewcommand\vec{\bm}

\newcommand{\uRe}{{\mathrm{Re}}\,}

\newcommand{\ud}{\,{\mathrm{d}}}

\def\uiiint{\int\!\!\!\int\!\!\!\int}
\DeclareMathOperator{\sinc}{sinc}

\def\ug{g}

\def\uHopf{{\cal{H}}} % Hopf index
\def\uvH{\vec{H}}     % -''-''- vector
\def\uK{K}         % anisotropy constant
\def\uvk{\vec{k}}    % incident beam wave vector
\def\uvkp{\vec{k}^\prime} % scattered neutron w-v
\newcommand{\uvM}{\vec{M}}
\newcommand{\uvm}{\vec{m}}
\newcommand{\umx}{m_{\mathrm{X}}}
\newcommand{\umy}{m_{\mathrm{Y}}}
\newcommand{\umz}{m_{\mathrm{Z}}}
\newcommand{\uqmx}{\widetilde{m}_{\mathrm{X}}}
\newcommand{\uqmy}{\widetilde{m}_{\mathrm{Y}}}
\newcommand{\uqmz}{\widetilde{m}_{\mathrm{Z}}}
\def\uO{O}         % Cartesian system origin

\newcommand{\uMs}{M_{\mathrm{S}}}
\newcommand{\uMssq}{M_{\mathrm{S}}^2}

\def\uvq{\vec{Q}}  % q vector
\def\uvnq{\vec{q}} % q vector magnitude\
\def\unq{q} % q vector magnitude\
\def\uqt{\theta_{\vec{q}}}
\def\uqf{\varphi_{\vec{q}}}
\def\uR{R}         % hopfion radius
\def\uvr{\vec{r}}          % radius-vector
\def\uvrho{\vec{\rho}}          % radius-vector
\def\ur{\rho}          % normalized radius-vector
\def\ut{\theta}
\def\uf{\varphi}
\def\uX{X}         % Cartesian X coordinate
\def\uY{Y}         % Cartesian Y coordinate
\def\uZ{Z}         % Cartesian Z coordinate
\def\uXp{\widetilde{X}} % Cartesian X primed
\def\uYp{\widetilde{Y}} % Cartesian Y primed
\def\uZp{\widetilde{Z}} % Cartesian Z primed
\graphicspath{ {./figures/} }

\def\uSigmad{\frac{\mathrm{d} \Sigma}{\mathrm{d} \Omega}}
\def\uSigmadInline{\mathrm{d} \Sigma/\mathrm{d} \Omega}

\def\uSigmadUU{\frac{\mathrm{d} \Sigma^{\uparrow\uparrow}}{\mathrm{d} \Omega}}
\def\uSigmadUUInline{\mathrm{d} \Sigma^{\uparrow\uparrow}/\mathrm{d} \Omega}
\def\uSigmadDD{\frac{\mathrm{d} \Sigma^{\downarrow\downarrow}}{\mathrm{d} \Omega}}
\def\uSigmadDDInline{\mathrm{d} \Sigma^{\downarrow\downarrow}/\mathrm{d} \Omega}
\def\uSigmadUD{\frac{\mathrm{d} \Sigma^{\uparrow\downarrow}}{\mathrm{d} \Omega}}
\def\uSigmadUDInline{\mathrm{d} \Sigma^{\uparrow\downarrow}/\mathrm{d} \Omega}
\def\uSigmadDU{\frac{\mathrm{d} \Sigma^{\downarrow\uparrow}}{\mathrm{d} \Omega}}
\def\uSigmadDUInline{\mathrm{d} \Sigma^{\downarrow\uparrow}/\mathrm{d} \Omega}
\def\uSigmadN{\frac{\mathrm{d} \Sigma_\mathrm{N}}{\mathrm{d} \Omega}}
\def\uSigmadNInline{\mathrm{d} \Sigma_\mathrm{N} / \mathrm{d} \Omega}
\def\uSigmadM{\frac{\mathrm{d} \Sigma_\mathrm{M}}{\mathrm{d} \Omega}}
\def\uSigmadMInline{\mathrm{d} \Sigma_\mathrm{M}/\mathrm{d} \Omega}
\def\uSigmadPerpM{\frac{\mathrm{d} \Sigma^\perp_\mathrm{M}}{\mathrm{d} \Omega}}
\def\uSigmadPerpMI{\frac{\mathrm{d} \Sigma^{\perp,\mathrm{I}}_\mathrm{M}}{\mathrm{d} \Omega}}
\def\uSigmadPerpMII{\frac{\mathrm{d} \Sigma^{\perp,\mathrm{II}}_\mathrm{M}}{\mathrm{d} \Omega}}

\def\uSigmadParM{\frac{\mathrm{d} \Sigma^\parallel_\mathrm{M}}{\mathrm{d} \Omega}}
\def\uSigmadParMIandII{\frac{\mathrm{d} \Sigma^{\parallel}_\mathrm{M}}{\mathrm{d} \Omega}}
\def\uSigmadParSFIandII{\frac{\mathrm{d} \Sigma^{\parallel}_\mathrm{SF}}{\mathrm{d} \Omega}}

\def\uSigmadSF{\frac{\mathrm{d} \Sigma_\mathrm{SF}}{\mathrm{d} \Omega}}
\def\uSigmadSFInline{\mathrm{d} \Sigma_\mathrm{SF}/\mathrm{d} \Omega}
\def\uSigmadPerpSF{\frac{\mathrm{d} \Sigma^\perp_\mathrm{SF}}{\mathrm{d} \Omega}}
\def\uSigmadPerpSFI{\frac{\mathrm{d} \Sigma^{\perp,\mathrm{I}}_\mathrm{SF}}{\mathrm{d} \Omega}}
\def\uSigmadPerpSFII{\frac{\mathrm{d} \Sigma^{\perp,\mathrm{II}}_\mathrm{SF}}{\mathrm{d} \Omega}}

\def\uSigmadParaSF{\frac{\mathrm{d} \Sigma^{\parallel}_{\mathrm{SF}}}{\mathrm{d} \Omega}}

\def\uChiral{\eta}

\def\uChiralPerp{\eta^\perp}
\def\uChiralPerpI{\eta^{\perp,\mathrm{I}}}
\def\uChiralPerpII{\eta^{\perp,\mathrm{II}}}

\begin{document}

\title{Small-angle neutron scattering signatures of magnetic hopfions}

\author{Konstantin L. Metlov}\email{metlov@donfti.ru}
\affiliation{Galkin Donetsk Institute for Physics and Engineering, R.~Luxembourg Street~72, 283048~Donetsk, Russian Federation}
\author{Andreas Michels}\email{andreas.michels@uni.lu}
\affiliation{Department of Physics and Materials Science, University of Luxembourg, 162A~Avenue de la Fa\"iencerie, L-1511 Luxembourg, Grand Duchy of Luxembourg}

\date{\today}

\begin{abstract}
Magnetic hopfions are three-dimensional localized magnetic topological solitons which can exist in the bulk of magnetic materials. Based on a Ritz model for magnetic hopfions in a chiral magnet, the unpolarized magnetic small-angle neutron scattering (SANS) cross section, the spin-flip scattering cross section, and the chiral function (characterizing the imbalance between the two spin-flip scattering amplitudes) are computed here analytically; while the real-space correlation function is obtained numerically. Features of these functions, specific to magnetic hopfions, are discussed. Our results enable the SANS method to be used for the detection of magnetic hopfions.
\end{abstract}


\maketitle

\section{Introduction}

Hopfions are three-dimensional topological objects, embedded within vector fields~\cite{Faddeev1975,Faddeev1976}. Magnetic hopfions exist in the vector field of the local magnetization of a magnetically ordered material~\cite{BK11-book,RKBDMB2022}. There is a substantial recent theoretical~\cite{lake2018,sutcliffe2018} and experimental~\cite{kent2021,altbir2023,zheng2023} progress in the search of magnetic hopfions (or parts of hopfions, such as torons) in restricted geometry, opening new prospects for spintronics applications. Yet, the true potential of hopfions can only be realized in a fully three-dimensional ``bulk'' setting.

To observe bulk hopfions, imaging of the magnetization vector field $\uvM(\uvr)$ inside a magnet with nanometer-scale resolution is required. One of such promising new techniques is magnetic nanotomography~\cite{CMSGHBRHCG2020}, but this method is strongly tied to the details of the x-ray absorption edge of a particular chemical element and requires substantial readjustment to be applicable to a wide range of magnetic materials.

Magnetic small-angle neutron scattering is another observational technique~\cite{Michels2021book}, which is able to penetrate the bulk of magnetic materials and which possesses the required nanometer-scale resolution. While SANS does not show the distribution of the magnetization vector directly (in real space), the scattering cross section as a function of the scattering vector $\uvq$ contains plenty of information on the magnetization vector in Fourier space.

In this work, based on a variational model of hopfions in a classical helimagnet~\cite{M2023_TwoTypes}, we compute (for two stable hopfion types) their SANS images in the perpendicular and parallel (when the incident wave vector of the neutron beam is perpendicular or parallel to the hopfion axis, coinciding with the external magnetic-field direction) scattering geometries. We also compute the chiral scattering function in the perpendicular geometry and analyze the corresponding real-space correlation functions.

\section{The hopfion model}

Our starting point is the model for a spherical hopfion~\cite{M2023_TwoTypes}, which is in turn based on the Whitehead ansatz~\cite{whitehead1947}. Consider a Cartesian coordinate system $\uvr=\{\uX,\uY,\uZ\}$ such that its $\uO\uZ$ axis coincides with the symmetry axis of the hopfion and the direction of the externally applied magnetic field $\uvH$. The normalized (by the saturation magnetization $\uMs = |\uvM|$) components of the local magnetization vector $\uvm=\uvM/\uMs=\{\umx,\umy,\umz\}$ inside the hopfion with the Hopf index $\uHopf=1$ can be expressed via a series of maps:
\begin{subequations}
\label{eq:model}
\begin{align}
\label{eq:stereogr}
&\{\umx+\imath\umy,\umz\}=\{2w, 1 - |w|^2\}/(1+|w|^2),\\
\label{eq:whitehead}
&w=e^{\imath\chi}u/v,\\
\label{eq:E3S3}
 &u=\frac{2(\uXp+\imath \uYp)R}{\uXp^2+\uYp^2+\uZp^2+R^2},
 \quad
 v=\frac{R^2-\uXp^2-\uYp^2-\uZp^2+\imath 2 \uZp R}{\uXp^2+\uYp^2+\uZp^2+R^2},
\\
\label{eq:physE3}
&\{\uXp,\uYp,\uZp\}=\frac{\{\uX,\uY,\uZ\}}{g(\sqrt{\uX^2+\uY^2+\uZ^2}/R)},
\end{align}
\end{subequations}
where $\chi$ is a parameter that allows one to specify the hopfion type, $\uR$ denotes the radius of the hopfion, and $\ug(\ur)$ is the hopfion profile function, such that $\ug(0)=1$, $\ug'(0)=0$, and $\ug(1)=0$. The intermediate variables are: the complex representation $w$ of the fixed-length $|\uvm|=1$ magnetization vector, whose end always lies on the unit sphere $S^2$; the complex coordinates $u$ and $v$ on the unit sphere $S^3$, such that $|u|^2+|v|^2=1$; and the coordinates in the intermediate extended Euclidean space $\{\uXp,\uYp,\uZp\}$. This representation defines the magnetization vector field $\uvm(\uvr)$, which depends on the two scalar parameters $\chi$ and $\uR$, and on the profile function $\ug(\ur)$. All the model assumptions are contained within Eq.~\eqref{eq:physE3}, which maps the physical space to the extended Euclidean space $\{\uXp,\uYp,\uZp\}$. The remainder of Eq.~\eqref{eq:model} applies equally well~\cite{M2024ms} to the (yet unknown for the classical helimagnet) exact hopfion solutions.

Based on the analysis of the classical micromagnetic energy functional (including isotropic exchange, Dzyaloshinskii-Moriya, Zeeman, uniaxial anisotropy, and magnetostatic energy terms), it can be shown~\cite{M2023_TwoTypes} that there are only two stable values of $\chi$, corresponding to the minima of the hopfion's total energy: $\chi=\pi/2$ and $\chi=3\pi/2$. These define two stable hopfion types, which we will consider here---type~I and type~II, respectively, shown in the upper inset of Fig.~\ref{fig:xsq}. Both types of $\uHopf=1$ hopfions are symmetric around the $\uO\uZ$ axis. The type~I hopfions contain a circular outer antivortex tube wrapped around a circular inner vortex tube. In the type~II hopfions the tube order is reversed.  The radii $\uR$ and profile functions $\ug(\ur)$ can be found by solving the corresponding Euler-Lagrange equation for the extremum of the total energy functional. This procedure is implemented as a supplemental Mathematica code in~\cite{M2023_TwoTypes} and as a Fortran program~\cite{magnhopf} in~\cite{M2024ms}.

While Eq.~\eqref{eq:model} does not involve the external magnetic field $\vec{H}$ explicitly, it can describe the magnetization process of the hopfion. The field enters the Euler-Lagrange equation~\cite{M2023_TwoTypes}, coupling it to the equilibrium hopfion profile $\ug(\ur)$ and radius $\uR$. Increase/decrease of the field magnitude causes redistribution of the magnetic moments inside the hopfion, making more/less of them aligned with the field direction. This dependence can be obtained numerically using the codes in~\cite{M2023_TwoTypes,M2024ms}. In a sense, each equilibrium hopfion profile $\ug(\ur)$ already has a certain external field and anisotropy ``imprinted'', along with other material constants.

In this work we focus on the qualitative features of the SANS observables and derive our results in normalized dimensionless coordinates $\uvrho=\uvr/\uR$ for an arbitrary hopfion profile $\ug(\ur)$. The specific examples (unless otherwise stated) will be given for $\ug(\ur)=1-\ur^2$, which satisfies all the boundary conditions and roughly resembles the profile of the equilibrium type~I hopfion in small magnetic fields and for negligible uniaxial anisotropy.

\section{SANS cross section and chiral function}

\begin{figure}[tb]
\begin{center}
\includegraphics[width=0.70\columnwidth]{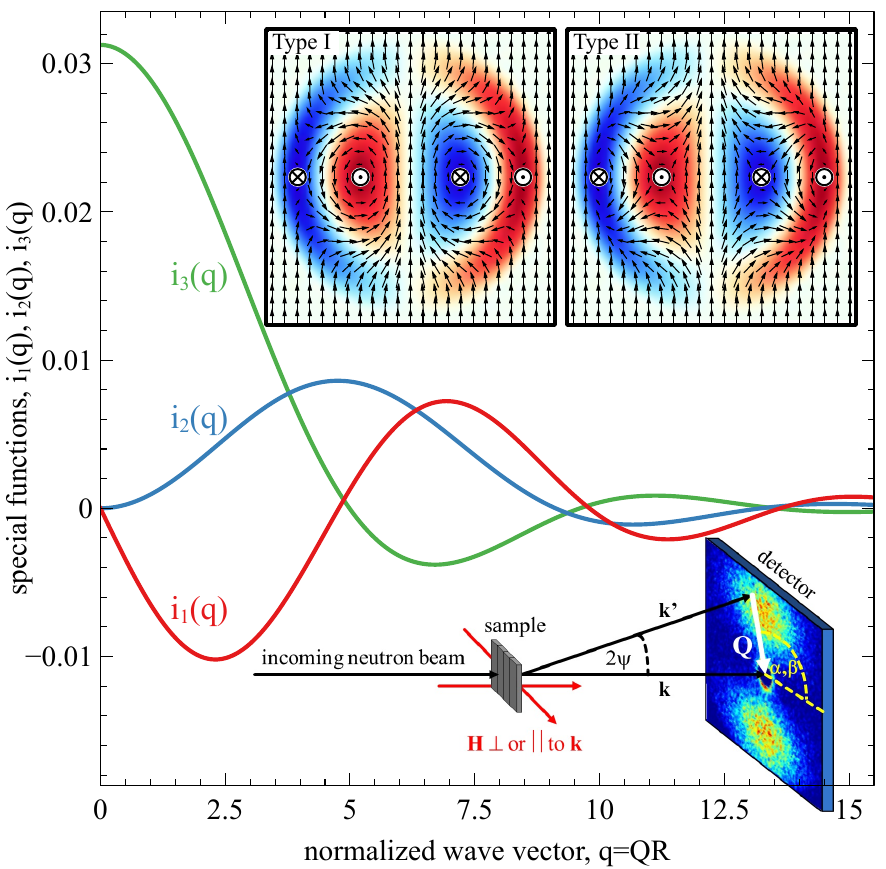}
\end{center}
\caption{\label{fig:xsq}Illustration of the special functions Eq.~\eqref{eq:specfun} (main plot). The upper inset shows the magnetization distribution Eq.~\eqref{eq:mReal} inside type~I and type~II hopfions for $\ug(\ur)=1-\ur^2$. The axis $\uO\uZ$ is vertical in the plane of the drawing. Note that the chirality of the type~II hopfion in the inset is reversed~\cite{M2023_TwoTypes} by reversing the sense of the magnetization circulation around the hopfion axis. This makes both hopfions in the inset correspond to a material with the same chirality (the sign of the Dzyaloshinskii-Moriya interaction constant). The lower inset depicts a sketch of the SANS setup.}
\end{figure}

In a neutron-scattering experiment (see the lower inset in Fig.~\ref{fig:xsq} for a basic sketch of the scattering geometry) a sample is irradiated by an approximately monochromatic neutron beam. An incident neutron with a wave vector $\uvk$ is scattered and acquires the wave vector $\uvkp$. It is then captured by a two-dimensional position-sensitive detector behind the sample, which counts how many neutrons have arrived into each of its pixels. These counts, after some renormalization and correction, yield the macroscopic differential scattering cross section $\uSigmadInline$ (usually expressed in units of \si{\per\centi\meter\per\steradian}), which is a function of the scattering vector $\uvq=\uvk-\uvkp$. In the small-angle regime we have $|\uvq|\ll|\uvk|$, and $\uvq$ is assumed to lie in the detector plane. For elastic scattering, the magnitude of the scattering vector is given by $Q = 4\pi/\lambda \sin\psi$, where $\lambda$ is the mean neutron wavelength and $2\psi$ denotes the scattering angle. The measurements are performed in
a magnetic field $\vec{H}$, coinciding in our case with the hopfion's axis of symmetry (see insets in Fig.~\ref{fig:xsq}) and the $\uO\uZ$ axis of the Cartesian coordinate system. The relative orientation of $\uvH$ and $\uvk$ defines a scattering geometry, of which two are most commonly used---the perpendicular ($\perp$) geometry, where the  magnetic field is applied perpendicular to the incident neutron beam ($\vec{H}\perp\uvk$), and the parallel ($\parallel$) geometry, where the field is along the incident beam ($\vec{H}\parallel\uvk$). It is convenient to parametrize the scattering vector by its magnitude and the polar angle in the perpendicular ($\uvq^\perp=\unq\{0,\sin\alpha,\cos\alpha\}/\uR$) and the parallel ($\uvq^\parallel=\unq\{\cos\beta, \sin\beta, 0\}/\uR$) scattering geometries, where $\unq=Q\uR$ is the dimensionless scattering vector magnitude.

While the neutrons (owing to their magnetic moment) are scattered by the magnetic moments in the sample, they are also scattered by the sample's nuclei. One of the main challenges in magnetic SANS is to separate these two contributions. This can e.g.\ be accomplished in a uniaxial polarization-analysis experiment~\cite{michels2010epjb,zakutna2020,ukleev2024}. In such an experiment, four separate scattering channels are considered, which are characterized by their respective cross sections:~$\uSigmadUUInline$, $\uSigmadDDInline$, $\uSigmadUDInline$, $\uSigmadDUInline$, where the arrows denote the two possible neutron spin~$1/2$ projections, before and after scattering, onto the direction of the externally applied magnetic (guide) field $\vec{H}$. In the present context, the following two combinations of $\uSigmadUDInline$ and $\uSigmadDUInline$ cross sections are particularly useful~\cite{maleyev2002}:
\begin{equation}
 \uSigmadSF = \frac{1}{2} \left(\uSigmadUD+\uSigmadDU\right)\,\,\,\mathrm{and}\,\,\,
 \uChiral = \frac{1}{2 \imath} \left(\uSigmadUD-\uSigmadDU\right).
\end{equation}
The first one of these ($\uSigmadSFInline$) is here simply denoted as the (polarization-independent) spin-flip SANS cross section; it is always positive and has the advantage that it does not contain the nuclear coherent SANS signal. The second term ($\uChiral$) is called the chiral function and it characterizes the nonreciprocity in the polarized neutron scattering cross section. Such a nonreciprocity can be expected in the presence of the Dzyaloshinskii-Moriya interaction (required for hopfion stability), which introduces a net chirality into the system.

Neutron polarization-analysis experiments are usually time-consuming and the related data reduction (spin-leakage correction) is challenging, so that often unpolarized measurements are carried out. When the SANS setup does not take the neutron polarization into account (no polarizing and analyzing optics), an unpolarized SANS cross section $\uSigmadInline$ is measured~\cite{maleyev2002}:
\begin{equation}
 \uSigmad = \frac{1}{2} \left( \uSigmadUU+\uSigmadDD+\uSigmadUD+\uSigmadDU \right) = \uSigmadN + \uSigmadM ,
\end{equation}
which consists of a nuclear ($\uSigmadNInline$) and a magnetic ($\uSigmadMInline$) part. We note that the nuclear SANS cross section $\uSigmadNInline$ does not depend (at least up to reasonably small values) on the magnitude of the external magnetic field, while the unpolarized magnetic SANS $\uSigmadMInline$ is usually strongly field-dependent. Moreover, for many systems, the $\uSigmadNInline$ adds as an isotropic ``background'' to the anisotropic $\uSigmadMInline$. Therefore, in view of this and since unpolarized experiments are faster and less complex than polarization-analysis experiments, we find it useful to also provide results for the unpolarized magnetic SANS cross section $\uSigmadMInline$.

In the following, we shall compute $\uSigmadMInline$, $\uSigmadSFInline$, and $\uChiral$ due to the scattering by hopfions for the perpendicular and parallel scattering geometries. The corresponding SANS cross sections can be expressed as~\cite{Michels2021book}:
\begin{subequations}
\label{eq:crossSectionsSrc}
\begin{align}
 \uSigmadPerpSF = & K \left(|\uqmx|^2 + |\uqmy|^2 \cos^4\alpha +|\uqmz|^2 \sin^2\alpha\cos^2\alpha - \uRe(\uqmy\overline{\uqmz}) \cos^2\alpha\sin2\alpha\right), \\
 \label{eq:sfpara}
 \uSigmadParaSF = & K \left(|\uqmx|^2 \sin^2\beta + |\uqmy|^2 \cos^2\beta - \uRe(\uqmx\overline{\uqmy}) \sin2\beta \right), \\
 \imath \uChiralPerp = &
 K \left((\uqmx\overline{\uqmy}-\overline{\uqmx}\uqmy)\cos^2\alpha - (\uqmx\overline{\uqmz}-\overline{\uqmx}\uqmz)\cos\alpha\sin\alpha\right), \\
  \uSigmadPerpM = & K \left(|\uqmx|^2 + |\uqmy|^2 \cos^2\alpha + |\uqmz|^2 \sin^2\alpha - \uRe(\uqmy\overline{\uqmz}) \sin2\alpha\right), \\
 \label{eq:unpolpara}
 \uSigmadParM = & K \left(|\uqmx|^2 \sin^2\beta + |\uqmy|^2 \cos^2\beta + |\uqmz|^2 - \uRe(\uqmx\overline{\uqmy})\sin2\beta\right),
\end{align}
\end{subequations}
where $K=8\pi^3\uMssq b_\mathrm{H}^2 V$, $b_\mathrm{H}=\SI{2.906e8}{\per\ampere\per\meter}$ is the magnetic scattering length (the value is given for the small-angle regime, where the atomic magnetic form factor can be approximated by 1), $V=4\pi R^3/3$ is the volume of the hopfion, a tilde over a quantity denotes its Fourier image
\begin{equation}
\label{eq:Fourier}
 \widetilde{A}(\uvq) = \frac{1}{(2\pi)^{3/2}V}\uiiint_{V}\!A(\uvr) e^{-\imath \uvq\cdot\uvr} \ud^3\uvr,
\end{equation}
so that $\{\uqmx(\uvq),\uqmy(\uvq),\uqmz(\uvq)\}$ are the Fourier images of the magnetization vector components, and the overline refers to the complex conjugate ($\overline{\imath}=-\imath$). It is assumed that $|\uvq|>0$ strictly, which allows us to ignore the constant magnetization background that only contributes at $|\uvq|=0$.  Note that the chiral function vanishes for the parallel geometry.

The next step is to compute the Fourier images $\{\uqmx, \uqmy, \uqmz\}$. In real space, the Cartesian magnetization vector components inside the hopfion can be expressed from the Eq.~\eqref{eq:model} as:
\begin{equation}
\label{eq:mReal}
\begin{bmatrix}
 \umx \\ \umy \\ \umz
\end{bmatrix}
=
\begin{bmatrix}
0 \\ 0 \\ 1 
\end{bmatrix}
-\frac{4 \ur \ug\sin\theta}{(\ur^2+\ug^2)^2}
\begin{bmatrix}
 \cos(\uf+\chi) & -\sin(\uf+\chi) & 0 \\
 \sin(\uf+\chi) &  \cos(\uf+\chi) & 0 \\
 0 &  0 & 1 \\
\end{bmatrix}
\begin{bmatrix}
 \ur^2-\ug^2 \\ 
 2 \ur \ug \cos\ut \\ 2 \ur \ug \sin\ut
\end{bmatrix},
\end{equation}
where we have used a spherical coordinate system $\{\uX,\uY,\uZ\}=\uR\ur\{\cos\uf\sin\ut, \sin\uf\sin\ut, \cos\ut\}$ and omitted the argument $\ug=\ug(\ur)$. Outside of the hopfion the magnetization is uniform and directed along the $\uO\uZ$ axis.

For computing the Fourier image of Eq.~\eqref{eq:mReal}, it is most convenient to take the Fourier integral [Eq.~\eqref{eq:Fourier}] in spherical coordinates (with volume element $R^3\ur^2\sin\ut\ud\ur\ud\ut\ud\uf$), while expanding the exponent in spherical harmonics:
\begin{equation}
 e^{-\imath\uvq\cdot\uvr} =
 4\pi\sum\limits_{l=0}^{\infty}
 \sum\limits_{m=-l}^{l}
 \imath^l j_l(\unq \ur)\overline{Y_l^m(\ut,\uf)}Y_l^m(\uqt,\uqf),
\end{equation}
where $j_l$ denotes the spherical Bessel function of the first kind, $Y_l^m$ are the spherical harmonics, $\uqf$ and $\uqt$ are the polar and azimuthal angles of the scattering vector $\uvnq=\uvq R$. Neglecting the uniform background [first term in Eq.~\eqref{eq:mReal}], which is irrelevant in SANS, we get:
\begin{equation}
\label{eq:mFourier}
\begin{bmatrix}
 \uqmx \\ \uqmy \\ \uqmz
\end{bmatrix}
=
%\begin{bmatrix}
%0 \\ 0 \\ B
%\end{bmatrix}
%\delta(\unq)
%+ 
\begin{bmatrix}
 -\imath\cos(\uqf+\chi) & -\sin(\uqf+\chi) & 0 \\
 -\imath\sin(\uqf+\chi) &  \cos(\uqf+\chi) & 0 \\
 0 &  0 & 1 \\
\end{bmatrix}
\left(
\begin{bmatrix}
i_1 \sin\uqt \\ i_2 \sin2\uqt \\ -i_2 \cos2\uqt
\end{bmatrix}
-\frac{1}{3}
\begin{bmatrix}
0 \\ 0 \\ i_2 + 4 i_3
\end{bmatrix}
\right),
\end{equation}
where the special functions $i_{1,2,3}=i_{1,2,3}(\unq)$ are given by:
\begin{align}
\label{eq:specfun}
\{i_1,i_2,i_3\} & =
\frac{3\sqrt{2}}{\pi^{3/2}}
 \int_0^1\frac{\{\ur^3 \ug (\ug^2-\ur^2)j_1(\unq\ur),\ur^4 \ug^2 j_2(\unq\ur),\ur^4 \ug^2 \sinc(\unq\ur)\}}{(\ur^2+\ug^2)^2} \ud\ur ,
\end{align}
with $\sinc x=\sin x/x$. The special functions depend on the hopfion profile $\ug(\ur)$ and for a simple illustrative case $\ug(\ur)=1-\ur^2$ are displayed in Fig.~\ref{fig:xsq}.
Substituting the Fourier images Eq.~\eqref{eq:mFourier} into Eq.~\eqref{eq:crossSectionsSrc} with $\chi=\pi/2$ and $\chi=3\pi/2$ for, respectively, type~I and type~II hopfions, we obtain:
\begin{subequations}
\label{eq:crossSections}
\begin{align}
 \label{eq:spinflipI}
 \frac{1}{K}\uSigmadPerpSFI & = \left(i_1^2 +\frac{4}{9}\left(i_2-2i_3\right)^2\cos^2\alpha\right)\sin^2\alpha, \\
 \label{eq:spinflipII}
 \frac{1}{K}\uSigmadPerpSFII & =  \left(i_1^2 +\frac{4}{9}\left(\left(2+3\cos2\alpha\right)i_2+2i_3\right)^2\cos^2\alpha\right)\sin^2\alpha, \\
 \label{eq:chiralI}
 \frac{1}{K}\uChiralPerpI & = -\frac{4}{3}i_1\left(i_2-2i_3\right) \cos\alpha\sin^2\alpha , \\
 \label{eq:chiralII}
 \frac{1}{K}\uChiralPerpII & = -\frac{4}{3}i_1\left(\left(2+3\cos2\alpha\right)i_2+2i_3\right) \cos\alpha\sin^2\alpha ,\\
  \label{eq:sigmaMperpI}
 \frac{1}{K}\uSigmadPerpMI & = \left(i_1^2 +\frac{4}{9}\left(i_2-2i_3\right)^2\right)\sin^2\alpha = \frac{1}{K}\uSigmadParMIandII \sin^2\alpha, \\
  \label{eq:sigmaMperpII}
 \frac{1}{K}\uSigmadPerpMII & =  \left(i_1^2 +\frac{4}{9}\left(\left(2+3\cos2\alpha\right)i_2+2i_3\right)^2\right)\sin^2\alpha, \\
 \label{eq:sigmaMpara}
 \frac{1}{K}\uSigmadParMIandII & = i_1^2 +\frac{4}{9}(i_2-2i_3)^2,\quad
 %\label{eq:spinflippara}
 \frac{1}{K}\uSigmadParSFIandII = i_1^2,
\end{align}
\end{subequations}
where the roman superscripts $\mathrm{I}$ and $\mathrm{II}$ label the respective type of the hopfion. The parallel cross sections [Eq.~\eqref{eq:sigmaMpara}] (both unpolarized and spin-flip) are independent of the hopfion type. Moreover, we note that the parallel spin-flip SANS signal differs from the unpolarized paralled SANS cross section only by a term $\propto |\uqmz|^2$ [compare Eq.~\eqref{eq:sfpara} and \eqref{eq:unpolpara}], which only affects its $q$~dependence. The chiral function for the parallel scattering geometry is zero. These expressions are analyzed in the next section.

\section{Results and discussion}

\begin{figure}[tb]
\begin{center}
\includegraphics[width=0.95\columnwidth]{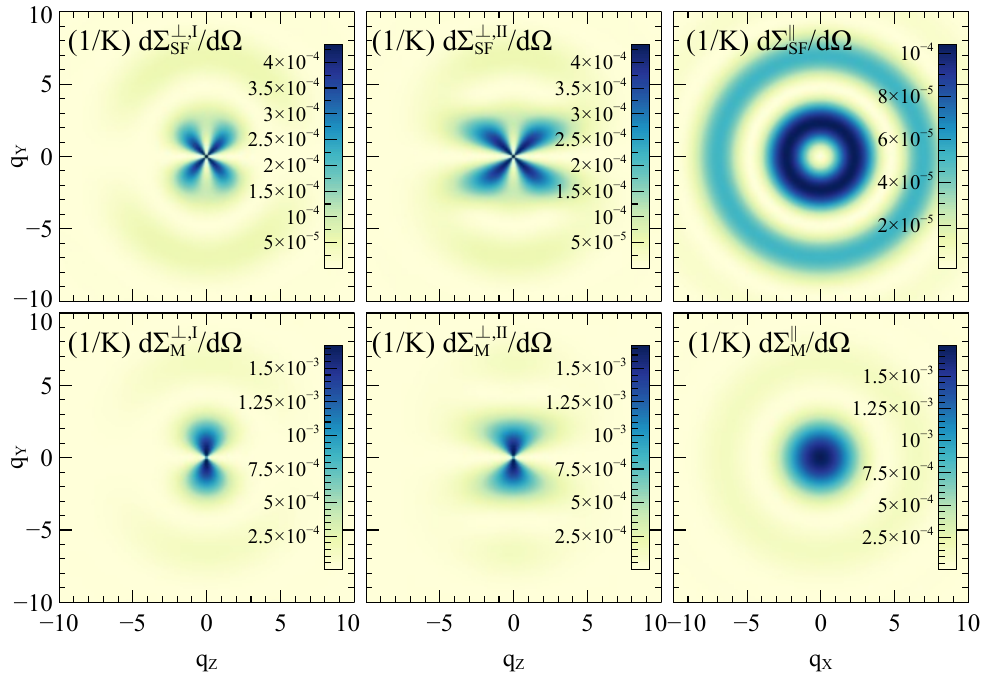}
\end{center}
\caption{\label{fig:S}Spin-flip and unpolarized perpendicular and parallel magnetic SANS cross sections [Eq.~\eqref{eq:spinflipI},  \eqref{eq:spinflipII}, \eqref{eq:sigmaMperpI}, \eqref{eq:sigmaMperpII}, \eqref{eq:sigmaMpara}] (in units of $\uK$) for type~I and type~II hopfions (see insets) with the profile function $\ug(\ur)=1-\ur^2$.}
\end{figure}

\begin{figure}[tb]
\begin{center}
\includegraphics[width=0.70\columnwidth]{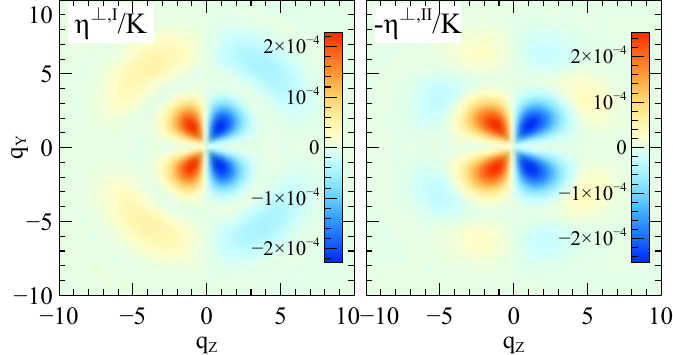}
\end{center}
\caption{\label{fig:chiperp} Chiral functions [Eq.~\eqref{eq:chiralI} and \eqref{eq:chiralII}] (in units of $\uK$) in the perpendicular scattering geometry for both hopfion types (see insets). Note that the chirality (the sign of $\uChiralPerp$) is reversed for type~II hopfions, as explained in Fig.~\ref{fig:xsq}.}
\end{figure}

The spin-flip and unpolarized SANS cross sections [Eq.~\eqref{eq:spinflipI},  \eqref{eq:spinflipII}, \eqref{eq:sigmaMperpI}, \eqref{eq:sigmaMperpII}, \eqref{eq:sigmaMpara}] for a particular simple $\ug(\ur)=1-\ur^2$ hopfion profile are plotted in Fig.~\ref{fig:S},
and the corresponding chiral functions [Eq.~\eqref{eq:chiralI} and \eqref{eq:chiralII}] are displayed in Fig.~\ref{fig:chiperp}.
While the hopfion profile $\ug(\ur)$ is integrated inside the special functions $i_{1,2,3}$ [compare Eq.~\eqref{eq:specfun}], in this way influencing their dependence on the magnitude $q$ of the scattering vector, the expressions~\eqref{eq:crossSections} are fully explicit with respect to the cross-section's angular ($\alpha$ or $\beta$) dependence.

The unpolarized and spin flip parallel SANS cross sections [Eq.~\eqref{eq:sigmaMpara}] are isotropic and hopfion-type independent. The perpendicular unpolarized cross section for type~I hopfions [Eq.~\eqref{eq:sigmaMperpI}] is particularly simple:~it equals the (isotropic) parallel cross section multiplied by a $\sin^2\alpha$~factor. Such an angular dependence is similar to the SANS cross section from a uniform sphere of reduced magnetization (relative to the constant background), but the radial dependence on $\unq$ is different (of course). The unpolarized perpendicular SANS cross section of type~II hopfions [Eq.~\eqref{eq:sigmaMperpII}] has additional terms with $\cos2\alpha\sin^2\alpha$ and $\cos^2 2\alpha\sin^2\alpha$ angular dependencies, which distinguish it from the type~I hopfion. The spin flip perpendicular cross sections [Eq.~\eqref{eq:spinflipI} and \eqref{eq:spinflipII}] generally display a dominating fourfold anisotropy visible in Fig.~\ref{fig:S}.

Unlike uniformly magnetized spheres, the hopfions are chiral, rendering their neutron-scattering response nonreciprocal, which results in a nonzero chiral function $\uChiralPerp$ and in an antisymmetric dependency on the momentum-transfer vector. The chiral functions for hopfions of both types are plotted in Fig.~\ref{fig:chiperp}; $\uChiralPerpI$ [Eq.~\eqref{eq:chiralI}] features a $\cos\alpha\sin^2\alpha$ angular anisotropy, while $\uChiralPerpII$ [Eq.~\eqref{eq:chiralII}] has an additional $\sin4\alpha\sin\alpha$ term.

It is interesting that from Eq.~\eqref{eq:crossSections} one can establish a universal relation between different spin-flip hopfion cross sections:
\begin{equation}
 4 \sin^2\alpha \uSigmadParaSF \uSigmadPerpSF = (\uChiralPerp)^2 + 4 \sin^4\alpha \left(\uSigmadParaSF\right)^2,
\end{equation}
which is equally valid for both hopfion types. It is expressed directly in terms of measurable quantities and does not involve explicitly the hopfion radius $\uR$ or its profile $\ug(\ur)$.

\begin{figure}[tb]
\begin{center}
\includegraphics[width=0.90\columnwidth]{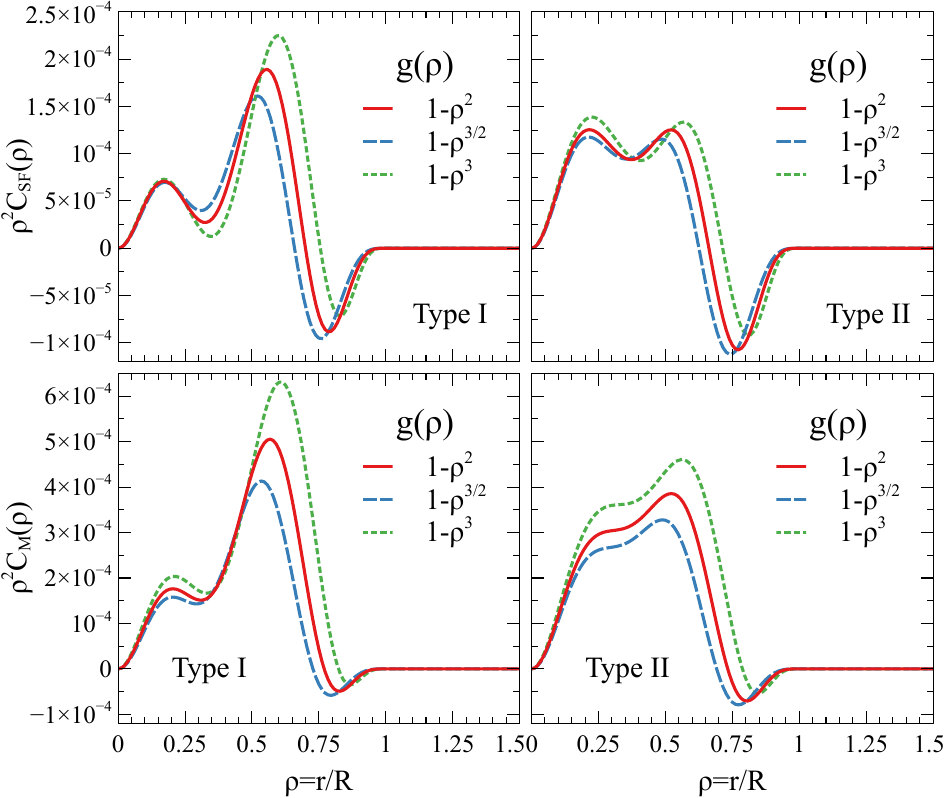}
\end{center}
\caption{\label{fig:corr}Correlation functions (in units of $\uK$) of the azimuthally-averaged unpolarized magnetic and spin-flip SANS cross sections in the perpendicular scattering geometry, $\ur^2 C_\mathrm{SF}(\ur)$ and $\ur^2 C_\mathrm{M}(\ur)$, respectively. The correlation functions are normalized by the spherical coordinate system volume element prefactor. Shown are results for hopfions of both types and for several hopfion profiles $\ug(\ur)$ (see insets).}
\end{figure}

Let us now consider more closely the radial dependence of the SANS cross section. A convenient tool for this is the real-space correlation function of the azimuthally-averaged cross section~\cite{Michels2021book,bersweiler2023}:
\begin{equation}
\label{eq:corrfunc}
C(\ur) = \int_0^\infty \left(\frac{1}{2\pi}\int_0^{2\pi} f(q,\alpha) \ud\alpha\right) \sinc(2 \unq \ur) \unq^2 \ud \unq ,
\end{equation}
where the function $f(q,\alpha)$ stands for either one of the expressions~\eqref{eq:crossSections}. The correlation function for the parallel unpolarized magnetic SANS cross section for both hopfion types follows $C_\mathrm{M}^{\parallel}(\ur)=2C_\mathrm{M}^{\perp,\mathrm{I}}(\ur)$; this can be directly seen from Eq.~\eqref{eq:sigmaMperpI}. The azimuthal average of the chiral function is zero, hence $C_\uChiral=0$. For the perpendicular cross sections, the correlation function Eq.~\eqref{eq:corrfunc} is evaluated numerically and the quantity $\ur^2 C(\ur)$ is plotted in Fig.~\ref{fig:corr} for both hopfion types and for several selected hopfion profiles.
Note that $\ur^2 C(\ur)$, compared to the pure $C(\ur)$, more directly reveals the spatial correlations, which will be otherwise masked by the growth of $C(\ur)$ towards $\ur=0$. This growth happens because $C(\ur)$ expresses the correlations in a spherical coordinate system, whose infinitesimal volume element scales as $r^2$.

One can see that all the dependencies in Fig.~\ref{fig:corr} contain, in general, a combination of two peaks. The peaks (regions of enhanced spin correlations) correspond to the magnetization outside of the vortex and antivortex tubes in a hopfion, while the minima between the peaks correspond to the tubes themselves, where the magnetization vector rotates fast. Note also that the constant magnetization background was neglected in our consideration (as it is usual in the SANS cross-section analysis), making the values of the correlation function zero at $\ur=0$ and for $\ur>1$. There are also zero-crossings in the correlation-function profiles, which were noted earlier in vortex-type spin structures~\cite{laura2020,bersweiler2023}.

\section{Conclusion}

For the relevant cases that the externally applied magnetic field (coinciding with the symmetry axis of the hopfion) is either perpendicular or parallel to the wave vector of the incident neutron beam, we have obtained analytical closed-form expressions for the unpolarized magnetic SANS cross section, the spin flip SANS cross section, and the chiral function of two types of stable hopfions. The latter can be used to characterize the nonreciprocity of the spin-flip scattering cross section. To analyze the dependence of the SANS cross sections on the magnitude of the scattering vector, we have computed the real-space correlation function, which exhibits two peaks, produced by vortex and antivortex tubes, and a zero crossing that are both characteristic for the hopfion texture. The two-peak structure clearly distinguishes hopfions from localized spherical inhomogeneities of the saturation magnetization. Qualitatively, a good indication of the hopfion scattering is a localized sphere-like scattering in an otherwise uniform material with a double-peak radial correlation function and a nonzero chiral function. Quantitatively, we establish a universal relation between different spin-flip SANS cross sections of hopfions. We hope that our results will make SANS instrumental in the ongoing search for bulk magnetic hopfions.

%\begin{acknowledgments}
%The support of the Russian Science Foundation under the project RSF~21-11-00325 is gratefully acknowledged.
%\end{acknowledgments}

%\bibliography{klm_base}
\end{document}